\documentclass[12pt]{article}
\usepackage{graphicx}% Include figure files
\newcommand{\sss}{\scriptscriptstyle}

\newcommand {\be}{\begin{equation}} % start equation
\newcommand{\ee}{\end{equation}}    % end equation

\def\dds1{\frac{\partial}{\partial s_1}}

\def\vtj{v_{{\sss T}j}}
\def\vti{v_{{\sss T}i}}

\def\vte{v_{{\sss T}e}}

\def\vtse{v_{{\sss T}se}}
\def\vtsi{v_{{\sss T}si}}
\def\vtsj{v_{{\sss T}sj}}
\def\vtfe{v_{{\sss T}fe}}
\def\vtfi{v_{{\sss T}fi}}
\def\vtfj{v_{{\sss T}fj}}

\def\d{d\kern-0.8 ex\vrule height 1.3 ex depth-1.24 ex width 0.7 ex
\kern 0.15 ex}
\def\D{D\kern-1.7 ex\vrule height .87 ex depth-0.8 ex width 0.7 ex
\kern 0.95 ex}

\evensidemargin
 0.3cm
\begin{document}
\baselineskip 20 pt

\begin{center}

\Large{\bf Kinetic instability of ion acoustic mode in  permeating
plasmas }

\end{center}

\vspace{0.7cm}

\begin{center}

{\bf  J. Vranjes and S. Poedts}

{\em K. U. Leuven, Center for Plasma Astrophysics, Celestijnenlaan
200B, 3001 Leuven,
 Belgium, and Leuven Mathematical Modeling and Computational Science Center
 (LMCC)}

\vspace{0.7cm}

{\bf Zahida Ehsan}

{\em Salam Chair and Physics Department, GC University Lahore 54000,
Pakistan.}

\end{center}

\vspace{2cm}

{\bf Abstract:} In plasmas with electron drift (current) relative to
static ions, the ion acoustic wave is subject to the kinetic
instability which takes place if the directed electron speed exceeds
the ion acoustic  speed. The instability threshold becomes different
in the case of one quasi-neutral electron-ion plasma propagating
through another static quasi-neutral (target) plasma. The threshold
velocity of the propagating plasma may be well below the ion
acoustic speed of the static plasma. Such a current-less instability
may frequently be expected in space and astrophysical plasmas.

\vspace{2cm}


\pagebreak

In an un-magnetized plasma with a macroscopic  velocity  of the
electron component $\vec v_{e0}=v_{e0} \vec e_z$ relative to  static
singly-charged ions, a kinetic instability of the ion acoustic mode
sets in, described  by the growth rate$^1$
\be
 \omega_i\! =\! \left(\frac{\pi}{8}\right)^{1/2} \!\!k c_s\left[\!
\left(\frac{m_e}{m_i}\right)^{1/2}\!\! \left(\frac{v_{e0}}{c_s}-
1\right) - \tau^{3/2} \exp\left(-
\frac{\tau}{2}\right)\!\right]\label{e1} \ee
Here, $c_s= (\kappa T_e/m_i)^{1/2}$ is the ion sound speed, and
$\tau=T_e/T_i$. The necessary condition
\be
v_{e0}> c_s[1+ (m_i/m_e)^{1/2} \tau^{3/2} \exp(-\tau/2)] \label{c1}
\ee
 may imply a rather
high electron current magnitude. A similar description of the ion
acoustic mode and the growth rate may be obtained also in the case
of a magnetized plasma$^2$ with the equilibrium magnetic field $\vec
B_0=b_0\vec e_z$ and for perturbations propagating in the direction
of the electron current $\vec v_{e0}=v_{e0} \vec e_z$ that is
parallel to the magnetic field vector. The instability described by
(\ref{e1}) is a purely kinetic, collision-less, electron inertia
effect. Its two-fluid counterpart$^{3,4}$ however is a strictly
electron-collision effect.

 The instability in (\ref{e1}) appears due to the Doppler shift term (the one containing $v_{e0}$), which may determine
 the sign of the growth rate.  However,  a much lower instability threshold may be obtained in the case of
two interpenetrating (permeating) plasmas. This is then a
{\em current-less}  instability, where  now the  Doppler shift  of ions appears to
play the main role.  This will be demonstrated in the forthcoming text.  Such  permeating plasmas can be created in lab conditions,
while in space this happens to be  a rather frequent  situation, for
example in the case of  colliding astrophysical clouds, in the
propagation of plasmas originating from the explosions of novae and
supernovae and moving through the surrounding plasmas, and also  in
the case of solar and stellar winds. In the case of the solar wind,
one electron-ion quasi-neutral component is generated mainly in the
polar regions (fast wind),  while the other electron-ion
quasi-neutral component (slow wind) originates from lower latitudes,
yet at large distances from the origin there may be overlapping
between the two. A most obvious example of such  permeating plasmas
is seen also in the solar atmosphere where plasma streams from lower
layers are  seen continuously propagating along magnetic field structures towards and through the solar corona.
 From the text and results that follow  it will become clear  that the present general study of two interpenetrating plasmas comprises
also, as special cases, various examples of plasmas containing some additional (electron or ion) species.

  In all of these examples we have one fast flowing
plasma whose parameters we shall denote by the subscript $f$, and
one  slow moving (target)  plasma  with the parameters  denoted by
the subscript $s$. In fact, the target plasma can be non-moving (or
'static', hence the same index 's'), or if it does move with some
constant velocity $\vec v_{s0}=v_{s0}\vec e_z$, the corresponding
equations for the whole combined plasma system  can conveniently be
written in the coordinate frame moving with this same velocity.
Therefore, in the forthcoming text the corresponding velocity of the
other component $\vec v_{f0}=v_{f0}\vec e_z$ will be used to
describe just the difference between the two.

In the case of singly-charged ions, the quasi-neutrality conditions
for the two plasmas in the equilibrium read:
\be
n_{fi0}=n_{fe0}=n_{f0}, \quad n_{si0}=n_{se0}=n_{s0}. \label{e2}
 \ee
In general $n_{s0}\neq n_{f0}$, and the same holds for the
temperatures of the two separate quasi-neutral plasmas $T_{fe}\neq
T_{se}$, $T_{fi}\neq T_{si}$.

We use the plasma distribution function for the species $j$,
\be
f_{j0}=\frac{n_{j0}}{(2\pi)^{3/2}\vtj^3} \exp\left\{-\frac{1}{2
\vtj^2}\left[v_x^2 + v_y^2 + (v_z- v_{j0})^2\right]\right\},
\label{e4} \ee
where $n_{j0}=const$, and $\vtj^2= \kappa T_j/m_j$. From the linearized
Boltzmann kinetic equation for the perturbed distribution function
\[
\frac{\partial f_{j1}}{\partial t} + \vec v \frac{\partial
f_{j1}}{\partial\vec r} + \frac{q_j }{m_j}\left( \vec
E_1 \frac{\partial f_{j0}}{\partial \vec v} +  \vec v\times
\vec B_0 \frac{\partial f_{j1}}{\partial \vec v}\right) = 0,
\]
and for longitudinal electrostatic perturbations $\sim \exp(-i
\omega + i k z)$ we obtain
\[
f_{j1}=- \frac{i q_j}{m_j} \frac{\vec E_1}{\omega - \vec k \vec v}\,
\frac{\partial f_{j0}}{\partial \vec v}.
\]
Further, using the Amp\`{e}re law
\[
0=\mu_0 \vec j+ \frac{1}{c^2} \frac{\partial \vec E}{\partial t},
\]
and the expression for the macroscopic current
\[
\vec j=\sum_j q_j \int \vec v f_j d^3 \vec v
\]
we arrive at
\be
1=\sum_j \frac{\omega_{pj}^2}{n_{j0} \omega \vtj^2}
\int_{-\infty}^{\infty} \frac{v_z(v_z-v_{j0})}{\omega - k v_z}
f_{j0} dv_x dv_y dv_z. \label{e5} \ee
The remaining integrations in (\ref{e5}) are straightforward and the
final result is
\be
1+ \sum_j \frac{1}{k^2 \lambda_{dj}^2} [1- Z(b_j)]=0. \label{e6} \ee
Here, $b_j=(\omega- k v_{j0})/(k \vtj)$,
$\lambda_{dj}=\vtj/\omega_{pj}$, $\omega_{pj}^2= q_j^2
n_{j0}/(\varepsilon_0 m_j)$, and
\[
Z(b_j)=\frac{b_j}{(2 \pi)^{1/2}} \int_c \frac{\exp(-
\zeta^2/2)}{b_j- \zeta} d\zeta
\]
is the plasma dispersion function, with the integration over the
Landau contour $c$. The dispersion equation (\ref{e6}) describes the
plasma (Langmuir) and the ion acoustic oscillations.

In the case $v_{sj0}=0$, $v_{fj0}=v_{f0}$,  expanding Eq. (\ref{e6})
in the ion acoustic frequency range on conditions
\be
k \vtsi \ll |\omega|\ll k \vtse, \quad |\omega - k v_{f0}|\ll k
\vtfe, \,\, k\vtfi, \label{c2} \ee
 where $ \vtsj^2=\kappa
T_{sj}/m_j$, $ \vtfj^2=\kappa T_{fj}/m_j$, we obtain the dispersion
equation for the ion acoustic wave
\[
\Delta(k, \omega)\equiv  1+ \frac{1}{k^2 \lambda_d^2} -
\frac{\omega_{psi}^2}{\omega^2} - \frac{3 k^2 \vtsi^2
\omega_{psi}^2}{\omega^4}
\]
\be
+ i \left(\frac{\pi}{2}\right)^{1/2} \left[\underbrace{\frac{\omega
\omega_{pse}^2}{k^3 \vtse^3}}_c + (\omega- k
v_{f0})\left(\underbrace {\frac{\omega_{pfe}^2}{k^3 \vtfe^3}}_a +
\underbrace{\frac{\omega_{pfi}^2}{k^3 \vtfi^3}}_b\right) +
\frac{\omega\omega_{psi}^2 }{k^3 \vtsi^3}\, \exp\left(-
\frac{\omega^2}{2 k^2 \vtsi^2}\right)\right]=0. \label{e8} \ee
Here, $1/\lambda_d^2=1/\lambda_{dse}^2 + 1/\lambda_{dfe}^2 +
1/\lambda_{dfi}^2$, and  $\lambda_{dse}=\vtse/\omega_{pse}$ etc. In
view of the conditions (\ref{c2}) the inertia  of the ion acoustic
(IA) mode is provided mainly by the ions of the static (target)
plasma.

In standard electron-ion plasma with streaming electrons and  for
$T_i\ll T_e$ it yields
\be
1+ \frac{1}{\lambda_{de}^2 k^2} - \frac{\omega_{pi}^2}{\omega^2} + i
(\pi/2)^{1/2} \left\{\frac{\omega_{pe}^2 (\omega- k v_{e0})}{k^3
\vte^3} + \frac{\omega_{pi}^2 \omega}{k^3 \vti^3} \exp\left[-
\omega^2/(2 k^2 \vti^2)\right]\right\}=0. \label{e7} \ee
From this one can further obtain the growth rate (\ref{e1}) and the
frequency $\omega^2 \simeq k^2 c_s^2$.
%\[
%\omega^2=\frac{k^2 c_s^2}{2} \left[1\pm (1+ 12
%T_i/T_e)^{1/2}\right]. \]

The only terms in (\ref{e8}) that may produce a growing mode are
those with the Doppler-shift, the two other terms in the imaginary
part always give the Landau damping. It is easily seen that the
electron part in the Doppler-shifted term is usually negligible
because  $a/b=(T_{fi}/T_{fe})^{3/2} (m_e/m_i)^{1/2}\ll 1$. Assuming
also that
\[
c/[b(\omega- k v_{f0})]=|\omega/(\omega- k v_{f0})| (n_{s0}/n_{f0})
(T_{fi}/T_{se})^{3/2} (m_e/m_i)^{1/2}\ll 1,
\]
 the  dispersion equation
becomes
\[
1+ \frac{1}{\lambda_d^2 k^2} - \frac{\omega_{psi}^2}{\omega^2} -
\frac{3 k^2 \vtsi^2 \omega_{psi}^2}{\omega^4}
\]
\be
 + i
(\pi/2)^{1/2} \left\{\frac{\omega_{pfi}^2 (\omega- k v_{f0})}{k^3
\vtfi^3} + \frac{\omega_{psi}^2 \omega}{k^3 \vtsi^3} \exp\left[-
\omega^2/(2 k^2 \vtsi^2)\right]\right\}=0. \label{e9} \ee
The frequency of the ion acoustic wave which follows from this is
\be
\omega_r^2=(k^2 \lambda_d^2 \omega_{psi}^2/2) [1 + (1+ 12
\lambda_{si}^2/\lambda_d^2)^{1/2}]. \label{fre} \ee
 For an ordinary electron-ion
plasma with $s$-species only, it takes the  shape $\omega_r^2=(k
c_s^2/2) [1 + (1+ 12 T_{si}/T_{se})^{1/2}]$.

The growth rate from (\ref{e9}),   $\omega_i\simeq - Im \Delta(k,
\omega_r)/[\partial (Re\Delta)/\partial \omega]_{\omega\simeq
\omega_r}$, becomes
\be
\omega_i=-\left(\frac{\pi}{8}\right)^{1/2} \frac{\omega_r^3}{k^2
\vtfi^2 (1+ 6 k^2 \vtsi^2/\omega_r^2)} \left\{\frac{\omega_r- k
v_{f0}}{k \vtfi} \frac{n_{f0}}{n_{s0}} + \frac{\omega_r}{k \vtsi}
\frac{T_{fi}}{T_{si}} \exp\left[-\omega_r^2/(2 k^2
\vtsi^2)\right]\right\}. \label{e11} \ee
In the limit $T_{fe}\gg T_{fi}$, $T_{se}\gg T_{si}$ and setting
$\omega_r\simeq k c_{se}$, the growth rate  becomes
\be
\omega_i=\left(\frac{\pi}{8}\right)^{1/2} k c_{se}
\left(\frac{T_{se}}{T_{fi}}\right)^{3/2}
\left\{\left(\frac{v_{f0}}{c_{se}}-1\right) \frac{n_{f0}}{n_{s0}}
-\left(\frac{T_{fi}}{T_{si}}\right)^{3/2} \exp\left[-T_{se}/(2
T_{si})\right]\right\}. \label{e10a} \ee
 From this the  instability threshold is
 \be
v_{f0}> c_{se} \left[1 + \frac{n_{s0}}{n_{f0}}
\left(\frac{T_{fi}}{T_{si}}\right)^{3/2} \exp\left(-\frac{T_{se}}{2
T_{si}}\right)\right]. \label{e10} \ee
Here, $c_{se}^2=\kappa T_{se}/m_i$. Eq. (\ref{e10}) has a remarkable
properties. It is seen that here [as compared to Eq. (\ref{c1})] the
first temperature ratio contains only the two {\em ion}
temperatures, and in addition  the electron-ion mass ratio vanishes.
In fact, the mass ratio can appear, yet in the present case it would
contain the {\em two ion masses} (that are taken as equal). For  the
singly charged two ion species, instead of $n_{s0}/n_{f0}$ in Eq.
(\ref{e10}) we would have $(n_{s0}/n_{f0})(m_{fi}/m_{si})$.
Consequently, in the present case the instability threshold may
become considerably lower. This is particularly valid for a flowing
plasma being accelerated into a less dense but hotter static target
plasma. One example of that kind may be  the plasma from the lower
solar atmosphere propagating through hot upper layers (e.g.,
spicules, and flows along  magnetic loops in general).

However, in general case the threshold velocity $v_{f0}$  for the
currentless instability may in fact be {\em below} the sound
velocity  of the target plasma.  As an example, the full Eq.
(\ref{e7}) is solved numerically by taking the wavelength
$\lambda=0.3$ m, and $T_{se}=10^5$ K, $T_{si}=10^3$ K,
$n_{f0}=10^{17}$ m$^{-3}$, and $n_{s0}$ is varied in the range shown
in Fig.~1.  The critical value of the flowing plasma velocity is
shown for several values of the temperature of the flowing plasma.
For the velocities above the given lines  the IA mode becomes
growing. For these parameters, in Eq. (\ref{fre}) the term
$\lambda_{si}^2/\lambda_d^2=0.07$ and also $\lambda_d \omega_{psi}=
0.38 c_{se}$. Therefore, contrary to (\ref{e1}) or (\ref{e10})  the
instability threshold (c.f. Fig.~1) is in fact well {\em below} the
IA speed  of the static (target) plasma. For lower $T_{fi}$ the
instability threshold is reduced because of a lower Landau damping
by the flowing ions.

\begin{figure}
\includegraphics[height=6.5cm, bb=14 14 295 225, clip=]{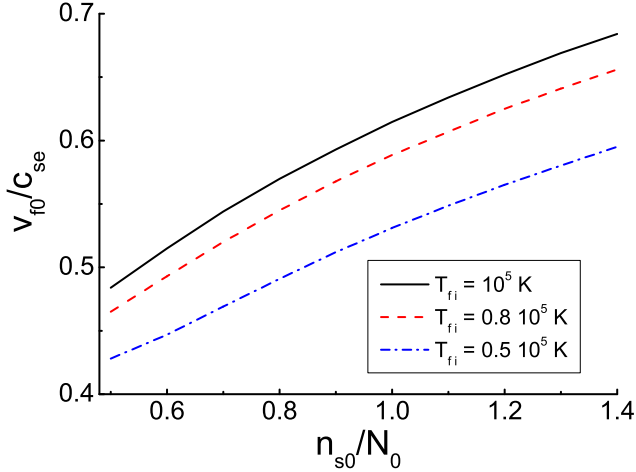}
\vspace*{-5mm} \caption{ The critical (threshold) values of the flowing
plasma velocity for the ion-acoustic wave  instability in terms of
the number density of the static (target) plasma. Here $c_{se}^2=
\kappa T_{se}/m_i$,  $N_0=10^{17}$ m$^{-3}$.
}\label{fig1}
% \vspace{0.3cm}
\end{figure}

Note that for the used parameters,  $a/b\ll 1$ but $c/b$ is on the
order of unity or larger. Yet, both the mode frequency and the
growth rate are well analytically described by Eqs. (\ref{fre}),
(\ref{e11}), respectively.

Instead of the conditions  (\ref{c2}), the expansion in Eq.
(\ref{e6}) could be done using  the following limits:
\be
k \vtsi \ll |\omega| \ll k \vtse, \quad  k \vtfi \ll |\omega- k
v_{f0}|\ll k \vtfe. \label{c3} \ee
The dispersion equation in this case becomes:
\[
\Delta(\omega, k) = 1+ \frac{1}{k^2 \lambda_{dse}^2} + \frac{1}{k^2
\lambda_{dfe}^2} - \frac{1}{k^2 \lambda_{dsi}^2} \frac{1}{b_{si}^2}
\left(1+ \frac{3}{b_{si}^2}\right) - \frac{1}{k^2 \lambda_{dfi}^2}
\frac{1}{b_{fi}^2} \left(1+ \frac{3}{b_{fi}^2}\right)
\]
\be
+ i \left(\frac{\pi}{2}\right)^2\left(\frac{b_{se}}{k^2
\lambda_{dse}^2} + \frac{b_{fe}}{k^2 \lambda_{dfe}^2}\right) + i
\left(\frac{\pi}{2}\right)^2\left[\frac{b_{si}}{k^2 \lambda_{dsi}^2}
\exp(-b_{si}^2/2) + \frac{b_{fi}}{k^2
\lambda_{dfi}^2}\exp(-b_{fi}^2/2)\right]=0. \label{e12} \ee
Here, $b_{fj}=(\omega- k v_{f0})/k \vtfj$, $b_{sj}=\omega/k \vtsj$.
 It is seen that as compared to Eq. (\ref{e9}), in the present case the mode is determined by both ion species, while the growth rate is an interplay between
   the fast ions and fast electrons.  The growth rate can  easily be discussed  in the limit
$T_{fi}=T_{si}$, $T_{se}=T_{fe}$ and for both plasmas with the same
density $n_0$. In this case the imaginary part of (\ref{e12})
becomes
\[
Im\Delta=\left(\frac{\pi}{2}\right)^2 \left\{
\frac{\omega_{pe}^2}{k^3 \vte^3} + \frac{\omega_{pi}^2}{k^3 \vti^3}
\exp\left[- \omega^2/(2 k^2 \vti^2)\right]\right\} (2\omega- k
v_{f0}).
\]
If in addition $\tau\equiv T_e/T_i\gg 1$, the ratio of the electron
and ion terms give the function $f(\tau)=(m_e/m_i)^{1/2} \tau^{-3/2}
\exp(\tau/2)$. For an electron-proton plasma we have $f(\tau)>1$ if
$\tau\geq 16$, and only in this limit  the growth rate is determined
by the electron flow term. Otherwise, the IA instability is driven
by the ion Doppler shift.

The cases of dominant ion (or electron) terms discussed above are
equivalent to studying also  electron-ion plasmas containing an
additional ion (or electron species).$^{5-8}$ As special cases,
these include also pair-ion,$^{9-12}$ negative-ion,$^{13}$
electron-positron-ion plasmas,$^{14,15}$ and also dusty plasmas
without$^{16}$ and with  oppositely charged grains.$^{17}$ Such
plasmas have been extensively studied in the recent past. The
general streaming kinetic instability presented above is directly
applicable to all these plasmas as special cases.

 \vspace{1cm}

\paragraph{Acknowledgements:}
The  results presented here  are  obtained in the framework of the
projects G.0304.07 (FWO-Vlaanderen), C~90347 (Prodex),  GOA/2009-009
(K.U.Leuven). Financial support by the European Commission through
the SOLAIRE Network (MTRN-CT-2006-035484) is gratefully
acknowledged.

\pagebreak

\end{document}